\documentclass{epl}

\title{Fine structure of the 0.7 MeV resonance in the $^{230}$Th \\
neutron--induced cross section\\
Short title: Fine structure in neutron-induced fission}
\author{M. Mirea\inst{1} \and L. Tassan-Got\inst{2} \and C. Stephan\inst{2}
\and C.O. Bacri\inst{2} \and R.C. Bobulescu\inst{3}}
\institute{
  \inst{1} Institute of Physics and Nuclear Engineering - P.O. Box MG-6, 
Bucharest, Romania\\
  \inst{2} Institut de Physique Nucl\'eaire - 91406 Orsay-Cedex, France\\
 \inst{3} Faculty of Physics, P.O. Box MG-11, Bucharest, Romania
}

\begin{document}

\maketitle

\begin{abstract}
The fine structure of the 0.7 MeV resonance in the $^{230}$Th neutron-induced
cross section is investigated within the hybrid model. A
very good agreement with experimental data is obtained.
It is suggested that
fine structure of the cross section quantify the
changes of the intrinsic states
 of the nucleus during the disintegration
process.
\end{abstract}

\section{Introduction}

The neutron-induced cross sections of $^{230,232}$Th exhibit 
multiple fine structures \cite{lyn,bol} superimposed on a gross structure of
the threshold cross section. If the fine structure is interpreted
as a series of rotational states constructed on a $\beta$-vibrational
state produced in some well of the multidimensional barrier, it
is straightforward to postulate the existence of a triple humped
barrier. The spacing between the members of the band is so small
that it is consistent only with a parent nucleus with prolate deformation
that reaches the vicinity of the second barrier top. Therefore, a shallow
minimum can be expected at this deformation to create a $\beta$-vibrational
state. Up to now, the assumption of a triple humped 
barrier seems to be the best 
interpretation for the fine structure of intermediate cross section resonances \cite{bol2}. 
The principal aim of the present work is to offer an alternative
explanation of this phenomenon by taking into account dynamical
single-particle effects. 

Recently, a Hybrid Model (HM) \cite{mir1} was developed in order to investigate
the intermediate structure of the fission cross section. In the frame
of the HM, the excited states during the deformation process of the parent
nucleus
and their realization probabilities
must be obtained. The occupations of the excited states are determined
theoretically by solving microscopic equations of motion.
These excited states are added to a phenomenological
double humped barrier and new barriers with different shapes are
constructed. The energy width in the fission channel
is proportional to the weighted summation of the penetrabilities of these barriers.
The fine structure of the 0.7 MeV resonance of the $^{230}$Th neutron-induced
cross section is studied within our model.

\section{Formalism}

Details concerning the HM can be found in ref. \cite{mir1} and only the main
steps that focuss strictly on the $^{230}$Th neutron-induced cross section
will be underlined in the following.

\begin{figure}
\onefigure{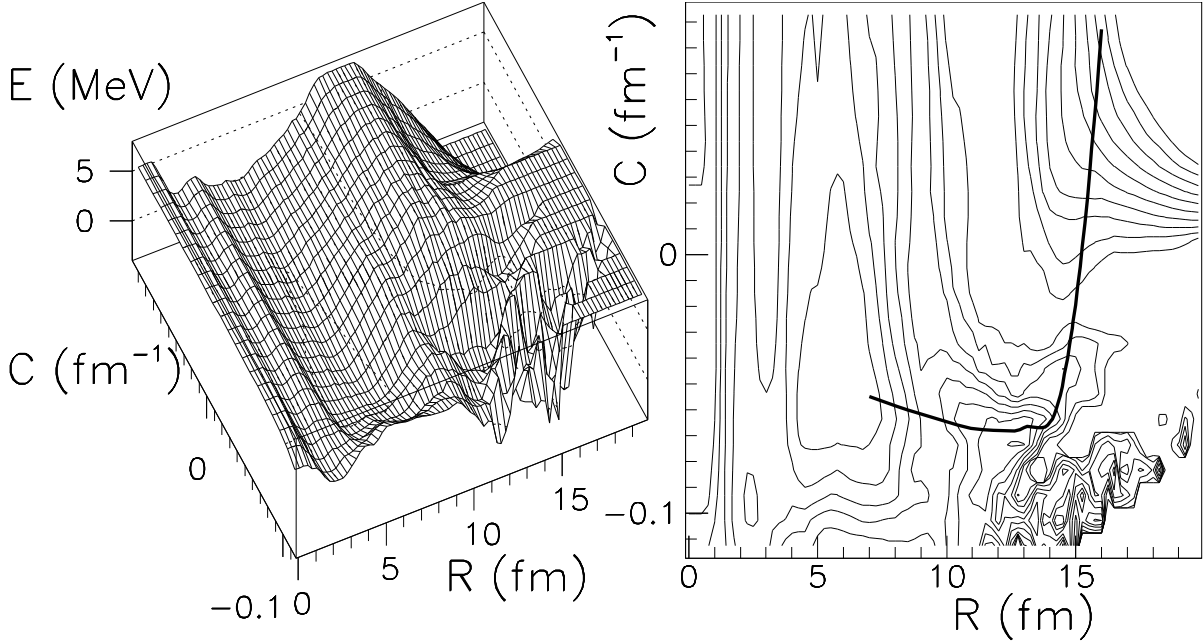}
\caption{Deformation energy in MeV for the partition
$^{231}$Th$\rightarrow^{100}$Zr+$^{131}$Sn. $C$ represents the curvature of the
neck and $R$ is the distance between the centers of the fragments.
The mass-asymmetry parameter is varied linearly with $R$ 
from 0 (in the ground-state configuration) up to
the final value (at the exit
point of the barrier).
In the right panel the step between two equipotential lines is 1 MeV.
The best action trajectory is represented with a thick line that starts
in the first well, penetrates the first barrier, attains the second
well and tunnels the second barrier towards scission. Positive
values of $C$ characterize necked-in shapes while negative values
correspond to neck-swollen shapes.}
\label{fig1}
\end{figure}

In order to determine the intrinsic single-particle states and their associated
occupation probabilities during the fission process it is necessary to 
perform a full calculation of the trajectory of the decaying system in the
available configuration space. For this purpose, 
an axial-symmetric nuclear shape
parametrization given by two spheres of different radii smoothly joined
by a neck region is used. This nuclear shape parametrization depends on the
most important macroscopic degrees of freedom encountered in fission,
namely elongation, necking and mass-asymmetry. The deformation energy is
obtained as a sum between the liquid drop energy and the shell effects.
The fission trajectory can be obtained by minimizing the action 
integral in the tridimensional configuration space. 
In fig. \ref{fig1}, the deformation energy for the
$^{231}$Th fission with heavy fragment $^{131}$Sn is displayed
together with the best trajectory. 
In this work, the liquid drop energy is
obtained in the frame of the Yukawa-plus-exponential model extended
for binary systems with different charge densities \cite{poe}. The shell effects
are determined using the Strutinsky prescriptions acting on level schemes
constructed with the Superasymmetric Two-Centre Shell Model (STCSM)\cite{mir2}. The
STCSM supplies single-particle states by adding correction terms
to a double oscillator eigenstates. In the STCSM, the most important
terms to be diagonalized are: the mass-asymmetry one, the neck one,
the spin-orbit interactions and the $l^{2}$ angular momentum interaction.
In order to perform the actual calculations the STCSM was drastically
improved. In the older version, the spin-orbit and $l^{2}$ operators were
constructed for a system given by two intersected spheres. In the present 
version, the spin interactions include a realistic dependence on the
neck formed between the nascent fragments.

\begin{figure}
\onefigure{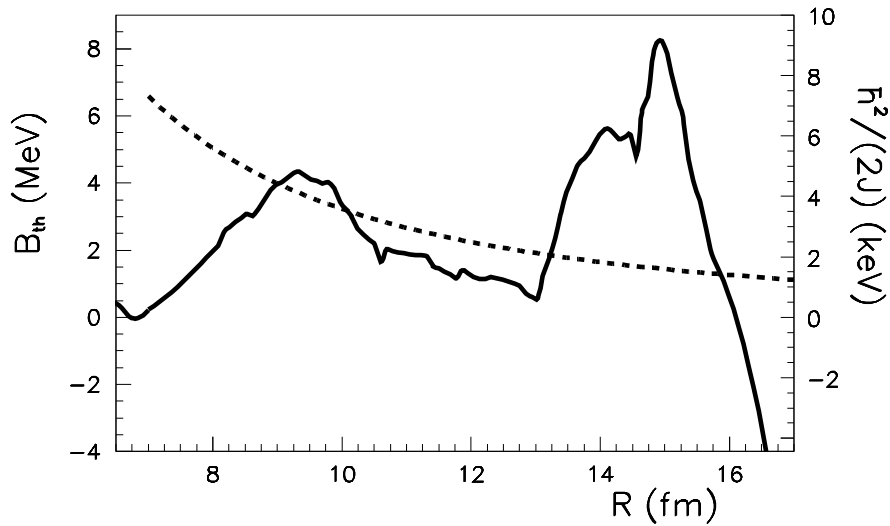}
\caption{Theoretical barrier along the minimal action trajectory as function
of the elongation. With
the dashed line the constant of inertia is plotted and the corresponding
values are displayed on the right scale. Values of
the moment of inertia constant close to typical ones are obtained for 
saddle configurations and wells.}
\label{fig2}
\end{figure}

The barrier obtained after the minimization procedure is plotted
in fig. \ref{fig2} together with the moment of inertia constant
$\hbar^{2}/(2J)$ where $J=(A_{0}/4)R^{2}$, $A_{0}$ being the mass
of the parent and $R$ is the distance between the centers of the
nascent fragments. The outer barrier is about three MeV lower in energy
in the actual version of the STCSM than in the previous one \cite{mir1,mir6}.
The barrier exhibits a narrow well in the
vicinity of the top of the second barrier. Our efforts are
focussed to explain the fine structure without appealing to the occurence
of this phenomenon.

\begin{figure}
\onefigure{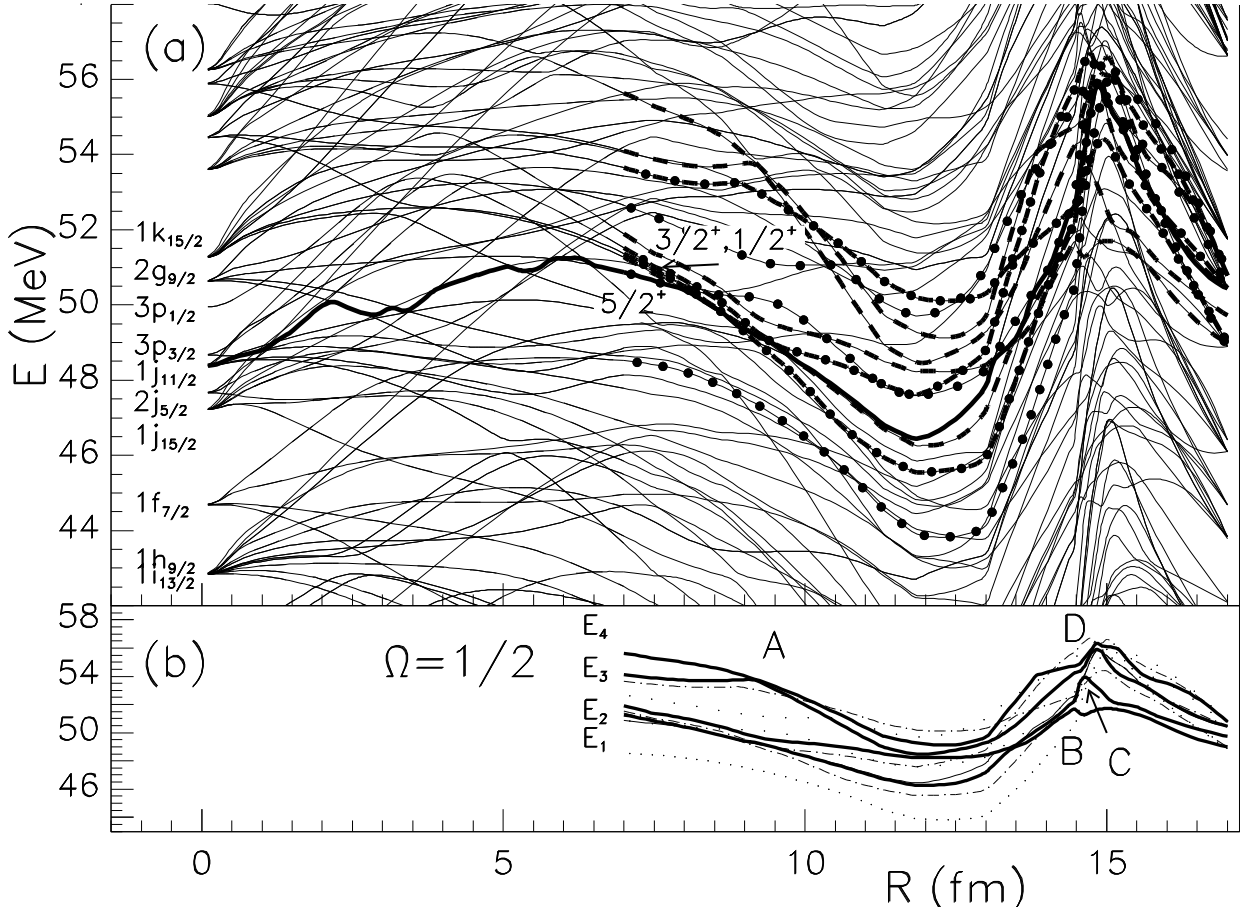}
\caption{(a) Neutron level scheme as function of the elongation. 
For zero elongation ($R$=0),
the shape parameterization describes a spherical nucleus. For low
values of the deformations, the system behaves as a Nilsson level
scheme. Asymptotically ($R\rightarrow\infty$), the two
level schemes of the formed fragments are superimposed.
The last occupied level is plotted with
a thick full line. Levels of interest are also plotted with different
thick line types: dashed line for four $\Omega$=1/2 levels, dashed-dotted line
for three $\Omega$=3/2 levels, dotted lines for two levels of spin projection
5/2 and one of 7/2. The Nilsson parameters $\kappa$ and $\mu$ are 0.0637 and 
0.74, respectively. (b) The four adiabatic $\Omega$=1/2 interesting levels are 
displayed with full thick
lines and numbered. The selected avoided
crossing regions between $\Omega$=1/2 levels are marked with letters. The 
other selected levels and the last occupied one are displayed with narrow
lines of same types as in plot (a).}
\label{fig3}
\end{figure}
The neutron single-particle level scheme is plotted in fig. \ref{fig3}(a).
For elongations comprised between $R$=0 (spherical shape) and
the $R\approx$7 fm (fundamental state) the shapes are considered
symmetrical by reflection, therefore the parity and the intrinsic 
spin projection $\Omega$
are good quantum numbers. From $R\approx$7 fm up to the scission,
the system loses the reflection symmetry to reach the final
partition with $^{131}$Sn heavy fragment and the parity is no longer
a good quantum number. Therefore, single-particle interactions can
be produced easily between levels characterized by the same value
of $\Omega$.
The spin interaction constants were choosed to reproduce as well as
possible the experimental sequence of the first excited levels
in the $^{231}$Th. Excepting the 5/2$^{-}$ bandhead, the first single-particle
excited states are retrieved for $R\approx$ 7 fm: 
5/2$^{+}$ (fundamental level), followed by
the 3/2$^{+}$ and 1/2$^{+}$ excited states. 
The fundamental state is considered in the present work to have
reflection-symmetry. If reflection-asymmetry or mass-asymmetry is
considered, the system acquires a doublet structure containing
both parities of $\Omega$, and the 5/2$^{-}$ bandhead will 
naturally appear in the sequence of levels. For the $^{231}$Th compound nucleus,
the lowest lying fission channels are essentially low-$\Omega$ single-particle states.  
Higher spins are not allowed by the compound nucleus formation cross section at
low energy.
Several single-particle levels that lie close to the
adiabatic last occupied level are selected as displayed in fig. \ref{fig3}(b).
The single-particle levels with the same good quantum numbers associated with some 
symmetries of the system cannot in general intersect but exhibit avoided level crossings 
\cite{sc,sc2}. Our system being characterized by an axial symmetry, the good quantum
numbers are the intrinsic spin projections $\Omega$. The radial coupling 
causes transitions of the
unpaired nucleon from one level to another of same $\Omega$. The probability
to jump from one level to another
can be evaluated by quantifying the Landau-Zener effect with a system
of microscopic equations of motion as realized in refs. \cite{mir3,mir4}. Solving
the coupled channel equations system,
the amplitudes of the single-particle wave functions in the nuclear level
distribution is determined.
In this way, the probability of occupation of an excited level is obtained.

For the four $\Omega$=1/2 levels, four avoided crossing regions are selected,
marked with letters A, B, C and D on the fig. \ref{fig3}(b). If the unpaired
nucleon is initially located on the level $E_{1}$, it can follow during
the disintegration any of the following paths 
$E_{2}BE_{1}$,
$E_{2}BCE_{2}$, $E_{2}BCDE_{3}$ or $E_{2}BCDE_{4}$
opened by the avoided 
crossing regions $B$, $C$ and $D$.
If the unpaired neutron
is initially located on another excited level, different energy paths
are open. 18 different excitation
channels can be obtained
within the selected configuration of only four levels and four avoided crossing regions. 
In the case of the three $\Omega$=3/2 selected levels, in
a similar manner, 13 different additional excitations are obtained.

\begin{figure}
\onefigure{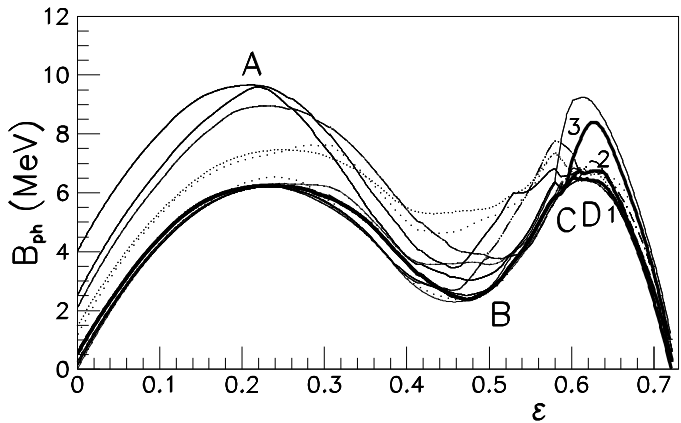}
\caption{Phenomenological barriers with single-particle excitations as
function of the dimensionless parameter $\epsilon$. The four avoided crossing
regions selected for $\Omega$=1/2 are marked with letters. The three excited
$\Omega$=1/2 selected barriers emerging from the adiabatic level
$E_{2}$ are plotted with full lines. These barriers are numbered and correspond to excitations
as explained in the text.
}
\label{fig4}
\end{figure}

The theoretical excitations computed in the frame of the superfluid
model are added to a phenomenological barrier. A
phenomenological barrier is conventionally simulated within three smoothed
joined parabolas \cite{cra}. The first barrier is labeled by A, 
the second one by B
and the second well by II. The mixing is realized in the most simplest way by
realizing a linear interpolation based on a correspondence between some
points ($R,\epsilon$) along the $x$-axis. $R$ is the elongation for
the theoretical model and $\epsilon$ the dimensionless parameter used in
simulating the phenomenological barrier. The correspondence was choosed for the
two minima, the two heights and the exit point. The hybrid model emerges.
New barriers are constructed as displayed in fig. \ref{fig4}. An imaginary
component of the potential is considered in the second well to take
into account other de-excitation channels apart the fission one. The
magnitude of this imaginary component increases with the excitation
energy of the compound nucleus and it is computed with the recipe of
ref. \cite{mir1}.
Finally, the barriers associated to the excitations and their weights are used
to determine the cross section by invoking the detailed balance principle.

\section{Results and discussion}

\begin{figure}
\onefigure{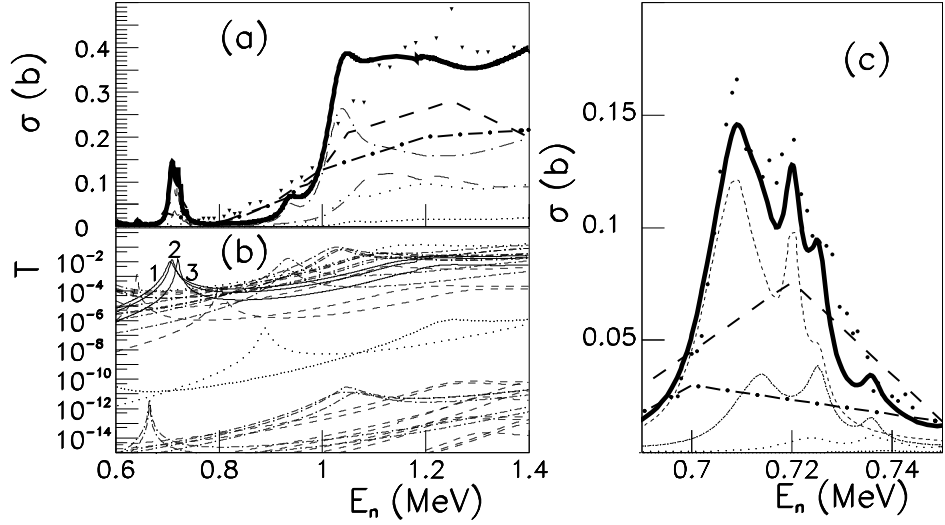}
\caption{(a) Cross section in the threshold region with respect
to the neutron energy determined in the
frame of HM compared with ENDF/B-VI.8 and JENDF-3.3 \cite{bnl} evaluations (thick dot-dashed and
dashed lines, respectively). Experimental
data extracted from ref. \cite{lyn} are also presented as down-point triangles.
The partial
contributions in the cross section of different spins of the compound nucleus 
are displayed with dashed lines for
$I$=1/2 ($I$ spin of the compound nucleus), 
dot-dashed lines for 3/2 and dotted lines for 5/2 and 7/2. (b) 
Weighted transmissions of the excited barriers that contribute to
the total cross section. Same line types
are used as in the plot (a) excepting the three transmissions corresponding
to barriers 1, 2 and 3 where a full line is used. (c) Detailed representation of the
cross section in the region 0.7 MeV. The spin dependent
partial cross sections are also given. The ENDF and JENDF evaluations
are plotted with dot-dashed and dashed lines, respectively. 
Experimental data from ref. \cite{bol}
are also given as full points.
}
\label{fig5}
\end{figure}
After a search of a set of parameters for the phenomenological barrier
consistent with experimental data,
the neutron-induced cross section is computed.
Results are plotted in fig. \ref{fig5} for
the following heights and stiffnesses of the phenomenological barrier:
$V_{A}$=6.24 MeV, $V_{II}$=2.37 MeV,  $V_{B}$=6.29 MeV,
 $\hbar\omega_{A}$=0.65 MeV, $\hbar\omega_{II}$=1 MeV and $\hbar\omega_{B}$=1.37 MeV. 
In the region [0.7,0.75] MeV a fine structure is found by the simulation
that
agrees very well with experimental data as evidenced in fig. \ref{fig5}(c). 
Unfortunately, the increasing
flank of the cross section around 1 MeV is not well reproduced. Other
parameters of the phenomenological barrier
succeed to reproduce better this region of the cross section
but in the same time lead to a degradation of the quality, for
reproducing the cross section
behavior around 0.7 MeV.
The main question concerns the origin of these fine structure resonances at 0.7 MeV.
This structure is produced by very close $\beta$-resonances of three excited 
single-particle
barriers of spin 1/2 superimposed on resonances due to rotations of the core.
The barriers with excitations created by the opened paths
$E_{2}BCE_{2}$, $E_{2}BCDE_{3}$ and $E_{2}BCDE_{4}$
corresponding to plot \ref{fig3}(b) are marked with 
the numbers 3, 1 and 2, respectively, on
fig. \ref{fig4}. 
In the vicinity of the top of the second barrier, (the elongation being
$R\approx$ 15 fm) a very large density of levels is revealed. This
behavior creates premises for an increased number of avoided crossing levels
regions
and,  therefore, for a similar number of small single-particle 
excitations. The unpaired neutron on the adiabatic level $E_{2}$
has a finite probability to jump on the 
level $E_{3}$ at point C and, successively,
on the level $E_{4}$ at point D (see fig. \ref{fig3}). 
Three excited barriers of interest can be obtained
in this way.  
The three excited barriers are different only in
a small interval around the top of the outer barrier up to the exit point, 
and therefore
give three different $\beta$-resonance very close in energy. These resonances
are plotted and marked in fig. \ref{fig5}(b) 
with their respective numbers 1, 2 and 3. 
These three resonances are in
an energy interval smaller than 0.05 MeV and produce three main fine
peaks in the fission cross section. Rotational resonances are
constructed on these $\Omega$=1/2 members as described in ref. \cite{mir1} leading to
additional structures. The structure due to rotations can be identified by
analyzing the fig. \ref{fig5}(c) where the role played by partial cross 
sections of spin $3/2$, $5/2$ and
$7/2$ can be acknowledged.
In the frame of HM, the members of a rotational band are produced by
$\beta$-resonances in a double barrier modified with 
a quantity $E_{r}(R,L)=L(L+1)\hbar^{2}/[2J(R)]$, where $J(R)$ is the 
moment of inertia
that depends on the elongation (see fig. \ref{fig2}) and $L$ the
orbital momentum of the core.

This investigation shows that the fine structure of the fission cross 
section can be explained by
the existence of several barriers associated to different single-particle
excitations. In the case of $^{230}$Th neutron-induced fission,
an unpaired neutron follows a well defined single-particle
adiabatic level during the deformation process and 
it is excited on other adiabatic
levels in the vicinity of the top of the second barrier, where the level 
density is very high. That causes the apparition of several peaks very close in energy
in the cross section. 
It can be assessed that dynamical single-particle effects can be
responsible for the fine structure phenomenon in the induced fission cross
section.

In the ref. \cite{bol2}, it was evidenced that a good fit of the 
0.7 MeV resonance behavior can be
realized only by taking into account two spin 1/2 bandhead resonances
created by slightly
different barrier parameters in the three humped picture. 
In the actual model, these bandheads
appears naturally and the variation of the barrier parameters are justified by the
existence of excitations close to the top of the barrier.
It can be also underlined that the actual evaluations of data files, as
evidenced by the large discrepancies revealed in fig. \ref{fig5}, fail
to reproduce the fine structure.

As a general conclusion, it is possible that the complex resonant 
structure of the fission cross section
is due to a rearrangement of orbitals and to the dynamics of the process, 
beginning
from the ground-state of the compound nucleus and reaching the scission.
A large number of different excited barriers are formed leading to a large
number of vibrational resonances in the second well. So, these resonances
carry information about the structure of the nucleus and the dynamics.



\end{document}